\documentclass[12 pt,onecolumn,reprint,amsmath,amssymb,aps,showpacs,eqsecnum,superscriptaddress,prx]{revtex4}

\usepackage{color,soul}
\usepackage[english]{babel}
\usepackage{graphicx}
\usepackage{bm}
\usepackage{epsfig}
\usepackage{amssymb}
\usepackage{amsfonts}
\usepackage{braket}
\usepackage{color}
\usepackage{epstopdf}
\usepackage{hyperref}
\usepackage{float}
\restylefloat{table}
\usepackage{bibentry}
\usepackage{color}
\usepackage{multirow}
\usepackage[caption=false]{subfig}
\usepackage{ulem}
\usepackage{lipsum}
\usepackage{blindtext}
\usepackage[export]{adjustbox}
\usepackage{mathrsfs}

\begin{document}
	\title{ Supersymmetry of $\mathcal{PT}$- symmetric tridiagonal Hamiltonians }

	\author{{\bf Mohammad Walid  AlMasri}}
	\email[Electronic address:$~~$]{mwalmasri2003@gmail.com}
	\affiliation{Department of Physics, Ko\c{c} University, Rumelifeneri Yolu, 34450 Sar\i yer, 
		Istanbul, Turkey}

	\begin{abstract} We extend the study of supersymmetric tridiagonal Hamiltonians to the case of  non-Hermitian Hamiltonians with real or complex conjugate eigenvalues. We find the relation between matrix elements of the non-Hermitian Hamiltonian $H$ and its supersymmetric partner $H^{+}$ in a given basis. Moreover, the orthogonal polynomials in the eigenstate
		expansion problem attached to $H^{+}$ can be recovered from  those polynomials arising
		from the same problem for $H$ with the help of kernel polynomials. Besides its generality, the developed formalism in this work is a natural home for using the numerically powerful Gauss quadrature techniques in probing the nature of some physical quantities such as the energy spectrum of $\mathcal{PT}$-symmetric complex potentials.  Finally,  we solve the shifted $\mathcal{PT}$-symmetric Morse oscillator exactly in the tridiagonal representation. 
	\end{abstract}

	\maketitle
	\section{Introduction}
	Supersymmetric quantum mechanics (SUSY-QM) plays an important rule in the study of exactly solvable potentials and its properties mainly due to its powerful factorization techniques which could be used in constructing new  analytically solvable potential  \cite{Witten,Krive,Haymaker,Cooper2001,Bagchi, Infeld, Dong}. The relation between one-dimensional solvable potentials and SUSY was addressed by Gendenshtein through  the concept of shape-invariant potentials \cite{Gend}. By definition the potential is said to be shape invariant if  its supersymmetric partner  has the same spatial dependence as the original potential with possibly adjusted parameters. \vskip 5mm
	It has been known for a  long time that unperturbed partial-wave kinetic energy operator $H^{0}= -\frac{1}{2} \frac{d^{2}}{dr^{2}}+ \frac{\ell (\ell+1)}{r^{2}}$ has a tridiagonal representation in a complete Slater or oscillator basis set $\{|\phi_{n}\rangle\}$ \cite{Fishman}. This  was extended later  to the case of Coulomb, Morse  and Rosen-Morse potentials \cite{haidari}. The  Supersymmetry of tridiagonal Hamiltonians was studied in \cite{Yamani}. 
	It was shown that if a positive semi-definite Hamiltonian has a tridiagonal representation in a given orthonormal basis $\{|\phi_{n}\rangle\}$, its supersymmetric partner Hamiltonian $H^{+}$ will have the same matrix representation in the same basis $\{|\phi_{n}\rangle\}$\cite{Yamani}. With the help of kernel polynomials, the orthogonal polynomials in the eigenstate
	expansion problem attached to $H^{+}$ can be recovered from  those polynomials arising
	from the same problem for $H$.  This method was elaborated to solve exactly the supersymmetry of Morse oscillator \cite{Yamani2} .\vskip 5mm
	
	 All previous studies were done with Hermitian tridiagonal Hamiltonians,  however
	it is known that  special class of non-Hermitian Hamiltonians may posses real eigenvalues \cite{Bessis,Bender,jones, Bender1, Bender2}. One feature of  these  Hamiltonians is that they are all $\mathcal{PT}$-symmetric,  note that the opposite of this statement is not true in general since not all $\mathcal{PT}$- symmetric Hamiltonians have real eigenvalues in their energy spectrum. These observations lead   to enormous studies of $\mathcal{PT}$-symmetric Hamiltonians and their properties especially in optics\cite{Optic,Musslimani,Ramezani,Longhi}. In \cite{Jin}, a  relation between  non-Hermitian systems with the Hermitian scattering systems was found. More recently Zhang et al.  proposed a  non-Hermitian supersymmetric array with high-order exceptional point of arbitrary order and investigated the topological properties of exceptional point \cite{Jin1}. Possible extension to the ideas of   $\mathcal{PT}$-symmetric Hamiltonian would be found in some magnetic systems such as quantum spin chains and  $\mathcal{PT}$-symmetric magnonic waveguides  \cite{Lin2,Berakdar1}.  The supersymmetric quantum mechanical treatment of $\mathcal{PT}$- symmetric Hamiltonians was done by Znojil et al \cite{Znojil1} and  the supersymmetric method for complex potentials was investigated in \cite{Mallik} and can be read from the general chapter by Levai  \cite{Levai3}.\\ 
	
	Throughout the present work, we  study the supersymmetry of tridiagonal non-Hermitian Hamiltonians with real or complex  conjugate pair eigenvalues. We formulate the problem using pseudo-Hermiticity notion  where the Hilbert space of quantum states is endowed with a Hermitian indefinite inner product. Very recently,  supersymmetry of tridiagonal non-Hermitian Hamiltonians was studied in \cite{tri} . However, our formalism is different from that presented in \cite{tri} since it  depends on the Askey scheme of classical  orthogonal polynomials. Also we considered cases where energy spectrum can have discrete and continuous values simultaneously unlike  \cite{tri} where the energy outcomes where discrete (shifted harmonic oscillator and squeezed Hamiltonian  ). The main motivation of this work is to develop the notion of $J$-matrix method for complex potentials and study the supersymmetry of these complex potentials in the tridiagonal representation.  Finally we tested the formalism by  solving  the $\mathcal{PT}$-symmetric Morse potential exactly in the tridiagonal representation domain with the help of orthogonal polynomials. \\ 
	The organization of paper goes as follows, in section (II) we give a formal definition of supersymmetric quantum mechanics and its pseudo-Hermitian generalization .  We briefly discuss the hierarchy method of $\mathcal{PT}$-symmetric tridiagonal Hamiltonians in section (III) . In section (IV), we express the non-Hermitian  Hamiltonians in tridiagonal representation and calculate  the matrix elements for $H$ and its superpartner $H^{(+)}$. The eigenstates expansion in terms of orthogonal polynomials are presented in (V).  Finally we apply  the developed formalism to the case of $\mathcal{PT}$-symmetric shifted Morse oscillator.    
	\section{Supersymmnetric Quantum Mechanics  and Pseudo-Hermiticity } Supersymmetry combines  the bosonic and fermionic degrees of freedom  in a unified algebra.  It was first proposed  in the context of unification of particle physics models\cite{Likhtman,Ramond,Wess}. However it was realized later that some ideas of supersymmetry could be used in solving nonrelativistic quantum mechanical problems\cite{Witten}. In SUSY-QM,   the  Hilbert space consists  of  bosonic $\mathcal{H}_{+}$ and fermionic $\mathcal{H}_{-}$ sub-Hilbert spaces    i.e.  $\mathcal{H}=\mathcal{H}_{+} \oplus\mathcal{H}_{-}$  for a given Hamiltonian $H$.

We define  the non-Hermitian supersymmetric  charges or generators  $Q^{I}$, $I=1,\dots N$ as 
	
	\begin{align}
		& Q^{I} (\mathcal{H}_{\pm})= \mathcal{H}_{\mp}, \\ 
		& Q^{I\; \dagger} (\mathcal{H}_{\pm})= \mathcal{H}_{\mp}, 
		\end{align}  
	where $\{ , \}$ is the anti-commutator. 
These generators satisfy the following super-algebra conditions 
	\begin{equation}\label{generators1}
		\{Q^{I}, Q^{J\; \dagger}\}= 2 \delta ^{IJ}\; H , 
		\end{equation}
		\begin{equation}\label{generators2}
		\{Q^{I}, Q^{I}\}=\{Q^{J\;\dagger}, Q^{J\; \dagger}\}=0, 
	\end{equation}
	where $I,J=1,\dots N$ and $\delta ^{IJ}$ is the Kronecker delta.\\
	 The  equations \ref{generators1} and \ref{generators2} can also be rephrased in the form 
	\begin{align}
		H=\frac{1}{2} (Q^{I}Q^{I\; \dagger}+ Q^{I\; \dagger}Q^{I}),\\ 
		(Q^{I})^{2}= (Q^{I\; \dagger})^{2}=0, 
	\end{align}
	For all $I=1, \dots N$. \\
	To simplify the formalism we write  the  $2N$-Hermitian SUSY charges as
	\begin{eqnarray}
	Q^{I}_{1}:= \frac{1}{\sqrt{2}} (Q^{I}+ Q^{I\; \dagger}) ,  \\ 
	Q^{I}_{2}:= \frac{1}{\sqrt{2}} (Q^{I}- Q^{I\; \dagger}). 
	\end{eqnarray}
	and the anti-commutators are 
	\begin{eqnarray}
	\{Q^{I}_{\alpha}, Q^{J}_{\beta}\}= 2 \delta_{\alpha \beta} \delta^{IJ} H, \\ 
	\{Q^{I}_{1}, Q^{I}_{2}\}=0 , 
	\end{eqnarray}
	where $\alpha, \beta= 1,2$ and $I,J= 1,2,\dots N$. Alternatively, the Hamiltonian takes the form 
	\begin{equation}
	H= (Q^{I}_{1})^{2}= (Q^{I}_{2})^{2}
	\end{equation}
In this work we concentrate on  $\mathcal{PT}$-symmetric SUSY-QM  \cite{Znojil1}.  By definition,  a $	\mathcal{PT}$-symmetric Hamiltonian satisfies 
	\begin{equation}
	\mathcal{PT} H (	\mathcal{PT})^{-1} = 	\mathcal{PT} H	\mathcal{PT}=H, 
	\end{equation}
	Where $\mathcal{P}$ and $\mathcal{T}$ are the parity and time-reversal operators respectively. These operators are defined in the following way 
	\begin{equation}
	\mathcal{P}\; x\;	\mathcal{P}=-x \; \; , \;  	\mathcal{P}\; p \;	\mathcal{P}=-p \; \; \; , 	\mathcal{T} i\; \mathbb{I} \mathcal{T}=-i \mathbb{I} .
	\end{equation}
	
	In the case of $\mathcal{PT}$-symmetric Hamiltonians,  the  inner product of two states $\phi_{1}$ and $\phi_{2}$ is defined  in the Hilbert space $\mathcal{H}$ as \cite{Bender}
	\begin{equation}\label{inner}
	\langle \langle \phi_{1}|\phi_{2} \rangle\rangle= \langle \phi_{1} | \mathcal{P} | \phi_{2}\rangle. 
	\end{equation} 
	which reduces to the ordinary Hermitian inner product when $\mathcal{P}=1$. In is worthy to note that $\{\mathcal{P}, Q\}=0$.  The inner product defined in \ref{inner} is invariant under time translation generated by the Hamiltonian $H$ if and only if $H$ is $\mathcal{P}$ pseudo-Hermitian as shown in  \cite{Mostafazadeh}. 
	The pseudo-superalgebra is given by \cite{Znojil1} ( we omitted the index $I$ for the sake of simplicity)
	\begin{eqnarray}
	\{Q, \tilde{Q}\}=2H,  \; \; \; Q^{2}=\tilde{Q^{ 2}}=0, 
	\end{eqnarray}
	Where $\tilde{Q}= \mathcal
	{P}^{-1} Q^{\dagger} \mathcal{P}$.\\
The Hamiltonian
	$H_{+}=  B A$ and $H_{-}=  A B$ can be  constructed from the shifting operator $A: \mathcal{H}_{+} \rightarrow \mathcal{H}_{-}$ which maps the eigenvectors of $H_{+}$ to those of $H_{-}$ while $B$ does the opposite action.

	\section{Factorization  of $\mathcal{PT}$-symmetric tridiagonal Hamiltonians}One advantage of supersymmetric quantum mechanics is the factorization techniques that allow us to factorize   Schrödinger equation by expressing the Hamiltonian as $H= A^{\dagger}A$+  constant term(s) \cite{Cooper2001, Sukumar,Levai3}. 
	The generic Hamiltonian with tridiagonal representation reads 
	\begin{equation}\label{general}
	H= -\frac{\hbar^{2}}{2m}\frac{d^{2}}{dr^{2}}+ \frac{\ell (\ell+1)}{2r^{2}}+ V_{1}(r)
	\end{equation} 
	so that $V^{eff}_{1}= \frac
{\hbar^{2}}{2m} \frac{\Psi_{0}^{\prime \prime}}{\Psi_{0}}$ 	where $'$ denotes the derivative with respect to the radial coordinate $r$ and  $V^{eff}_{1}= \frac{\ell (\ell+1)}{2 r^{2}}+ V_{1}$ is the centrifugal effective potential which appears usually in three-dimensional quantum problems with spherical symmetry. 
It can be defined using the superpotential $W=- \frac{\hbar}{\sqrt{2m}} [\frac{d\mathrm{ln}\Psi_{0}
	}{dr}]$ as \cite{Cooper2001}
	\begin{equation}
	V^{eff}_{1}= W^{2}- \frac{\hbar}{\sqrt{2m}} \frac{dW}{dr}
	\end{equation}

The ground state wavefunction can be computed from the superpotential up to a normalization constant as 
	\begin{equation}
	\Psi_{0} \sim exp[-\frac{\sqrt{2m}}{\hbar} \int ^{r} W(r^{\prime}) dr^{\prime}]
	\end{equation}
	 The Hamiltonian \ref{general} factorizes  as 
	\begin{align}
		H= A^{\dagger}A \\
		H^{\dagger}= A A^{\dagger} 
	\end{align}
	Where the bosonic operator $A$ and its Hermitian conjugate are 
	\begin{eqnarray}
	A= \frac{\hbar}{\sqrt{2m}}\frac{d}{dr}+W(r),\; \;  A^{\dagger}= -\frac{\hbar}{\sqrt{2m}}\frac{d}{dr}+W(r).
	\end{eqnarray}
	For unbroken supersymmetry, the ground state energy  is zero. Thus , we have 
	\begin{equation}\label{ground}
	H_{1}\Psi_{0}= \frac{-\hbar^{2}}{2m} \frac{d^{2}\Psi_{0}}{dr^{2}}+ \frac{\ell (\ell+1)}{2 r^{2}}\Psi_{0}+ V_{1}\Psi_{0}=0
	\end{equation}
	We can solve the  differential equation \ref{ground} and find an expression for $V^{eff}_{1}$ defined with respect to the ground state wave function and its second derivative with respect to the radial coordinate $r$,
	\begin{eqnarray}
	V^{eff}_{1}= \frac{\hbar^{2}}{2m } \frac{\Psi^{''}_{0}}{\Psi_{0}}
	\end{eqnarray} 
 The Hamiltonian \ref{ground} factorizes as 
	\begin{eqnarray}
	H_{1}= -\frac{\hbar^{2}}{2m} \frac{d^{2}}{dr^{2}}+ V_{1}^{eff}(r)\\ 
	= A_{1}A^{\dagger}_{1}+ E^{(0)}_{1}. 
	\end{eqnarray}
	After some calculations we get the Riccati equation for the superpotential $W_{1}$, 
	\begin{equation}
	W^{2}_{1}- W^{\prime}_{1}= \frac{2m}{\hbar^{2}}(V^{eff}_{1}(r)- E^{0}_{1}). 
	\end{equation} 
	The supersymmetric partner of Hamiltonian $H_{1}$ is 
	\begin{align}
		H_{2}= -\frac{\hbar^{2}}{2m} \frac{d^{2}}{dr^{2}}+ V_{2}^{eff}(r)\\= A^{\dagger}_{2} A_{2}+ E^{(0)}_{2}.
	\end{align}
	and the corresponding Riccati equation takes the form 
	\begin{equation}
	W^{2}_{2}+ W^{\prime}_{2}= \frac{2m}{\hbar^{2}}(V^{eff}_{2}(r)- E^{(0)}_{2}). 
	\end{equation}
	
	By iteration we can express the general Hamiltonian and Riccati equations respectively as 
	\begin{eqnarray}
	H_{n}= -\frac{\hbar^{2}}{2m} \frac{d^{2}}{dr^{2}}+ V_{n}^{tot}(r)\\ \nonumber 
	= A_{n}A^{\dagger}_{n}+ E^{(0)}_{n}, \; \;  n=1,3,5\dots \\ 
		H_{n}= -\frac{\hbar^{2}}{2m} \frac{d^{2}}{dr^{2}}+ V_{n}^{eff}(r)\\ \nonumber
	= A^{\dagger}_{n} A_{n}+ E^{(0)}_{n}, \; \;  n=2,4,6\dots 
	\end{eqnarray}
	\begin{eqnarray}
	W^{2}_{n}- W^{\prime}_{n}= \frac{2m}{\hbar^{2}}(V^{tot}_{n}(r)- E^{0}_{n}), n=1,3,5\dots  \\ 
	W^{2}_{n}+ W^{\prime}_{n}= \frac{2m}{\hbar^{2}}(V^{eff}_{n}(r)- E^{0}_{n}), n=2,4,6\dots  
	\end{eqnarray} 
	For unbroken supersymmetry, the partner Hamiltonian obey the following expressions 
	\begin{equation}
	E^{0}_{n+1}= E^{1}_{n}
	\end{equation}
	with $E^{0}_{0}=0$ and $n=0,1,2,\dots$. 
	
It is not difficult to notice that such procedure should be modified  for complex potentials with $\mathcal{PT}$-symmetry since $V^{eff}$ and $W$ are complex and should be decomposed to its real and imaginary parts \cite{Mallik}.\\
In order to factorize any  $\mathcal{PT}$-symmetric Hamiltonians,  we  introduce the following four related operators \cite{Znojil1}
\begin{eqnarray}
A^{(\pm)}= \frac{\hbar}{\sqrt{2m}}\frac{d}{dx}+ W^{(\pm)}(x)\; \; \;, \; B^{(\pm)}= -\frac{\hbar}{\sqrt{2m}}\frac{d}{dx}+W^{(\pm)}(x)
\end{eqnarray}
and the corresponding Riccati equation is 
\begin{equation}
H^{(\pm)}= B^{(\pm)}A^{(\pm)}= -\frac{\hbar^{2}}{2m}\frac{d^{2}}{dx^{2}}+ [W^{(\pm)}(x)]^{2}-\frac{\hbar}{\sqrt{2m}}[W^{(\pm)}(x)]^{\prime}
\end{equation}
where  $\prime$  is  the derivative with respect to the coordinate $x$. In the incoming sections we will choose the $+$ operators only since the second case follows the same procedures.  
 The term $ \frac{\ell (\ell+1)}{2 r^{2}}$  appears in the three-dimensional problems such as the    radial Coulomb potential and isotropic harmonic oscillator \cite{Haymaker}. 
 \section{J-matrix method for complex tridiagonal Hamiltonians }
 In ordinary quantum mechanics, it is customary to expand the Hamiltonian in a basis
 that makes it diagonal. This simplifies the process of computing the energy eigenvalues
 especially for large-sized Hamiltonian matrices. On the other hand, this approach reduces
 the class of possible solutions of the corresponding wave equation of the system. One possible extension of the space of solutions is known as J-matrix method which assumes the
 matrix representation of the Hamiltonian to be tridiagonal in a given orthogonal polynomial basis  \cite{Fishman,haidari,Mourad}. In the J-matrix method, we expand the wavefunctions over a complete set of
square-integrable functions $L^{2}(\mathbb{R})$ in configuration space $ |\psi(E,x)\rangle= \sum_{n} f_{n}(E) |\phi_{n}(x)\rangle$. The
 coefficients are some functions of energy. They encode all physical information about the
 system. The wave equation $H|\psi\rangle=E|\psi\rangle$ becomes 
 \begin{equation}
 \sum_{n} f_{n} H|\psi\rangle = E \sum_{n} f_{n} |\phi_{n} \rangle
 \end{equation}
 By projecting $\langle \phi_{m}$ from left we find 
 \begin{equation}
 \sum_{n} f_{n} \langle \phi_{m}|H|\phi_{n}\rangle = E \sum_{n} f_{n} \langle \phi_{m}|\phi_{n}\rangle 
 \end{equation}
which can written in matrix from as  $\sum_{n} H_{mn} f_{n} =E\sum_{n} \Omega_{mn}f_{n}$ where $\Omega_{mn}= \langle \phi_{m}|\phi_{n}\rangle $ is the overlap identity matrix of the basis elements. Our main task in this approach is to calculate
 the matrix elements of the tridiagonal and symmetric operator $J_{mn}$
 \begin{equation}\label{j}
 J_{mn}= H_{mn}- E \Omega_{mn} = (a_{n}-z) \delta_{n,m}+ b_{n}\delta_{n,m-1}+ b_{n-1} \delta_{n,m+1}
 \end{equation}
 In this work we extend J-matrix method to complex potentials with $\mathcal{PT}$ symmetry. For this
 reason we assume the matrix $J_{mn}$ to be non-self adjoint so that the relation  \ref{j}   will be written generally as 
 \begin{equation}
 J_{mn}= H_{mn}- \varepsilon \Omega_{mn} = (a_{n}-z) \delta_{n,m}+ b^{\star}_{n}\delta_{n,m-1}+ b_{n-1} \delta_{n,m+1}
 \end{equation}
 where in this case  the coefficients $a_{n}$ and $b_{n}$ are complex quantities in general. For a given
 non-self adjoint Hamiltonian with one-dimensional complex potential $V$, the reality of its
 energy bound states requires $[\mathcal{PT},H]=0$ and $(V(-x))^{\star}=V(x)$ \cite{Levai1}

\begin{figure}\label{sin}
	\centering
	\includegraphics[width=12cm]{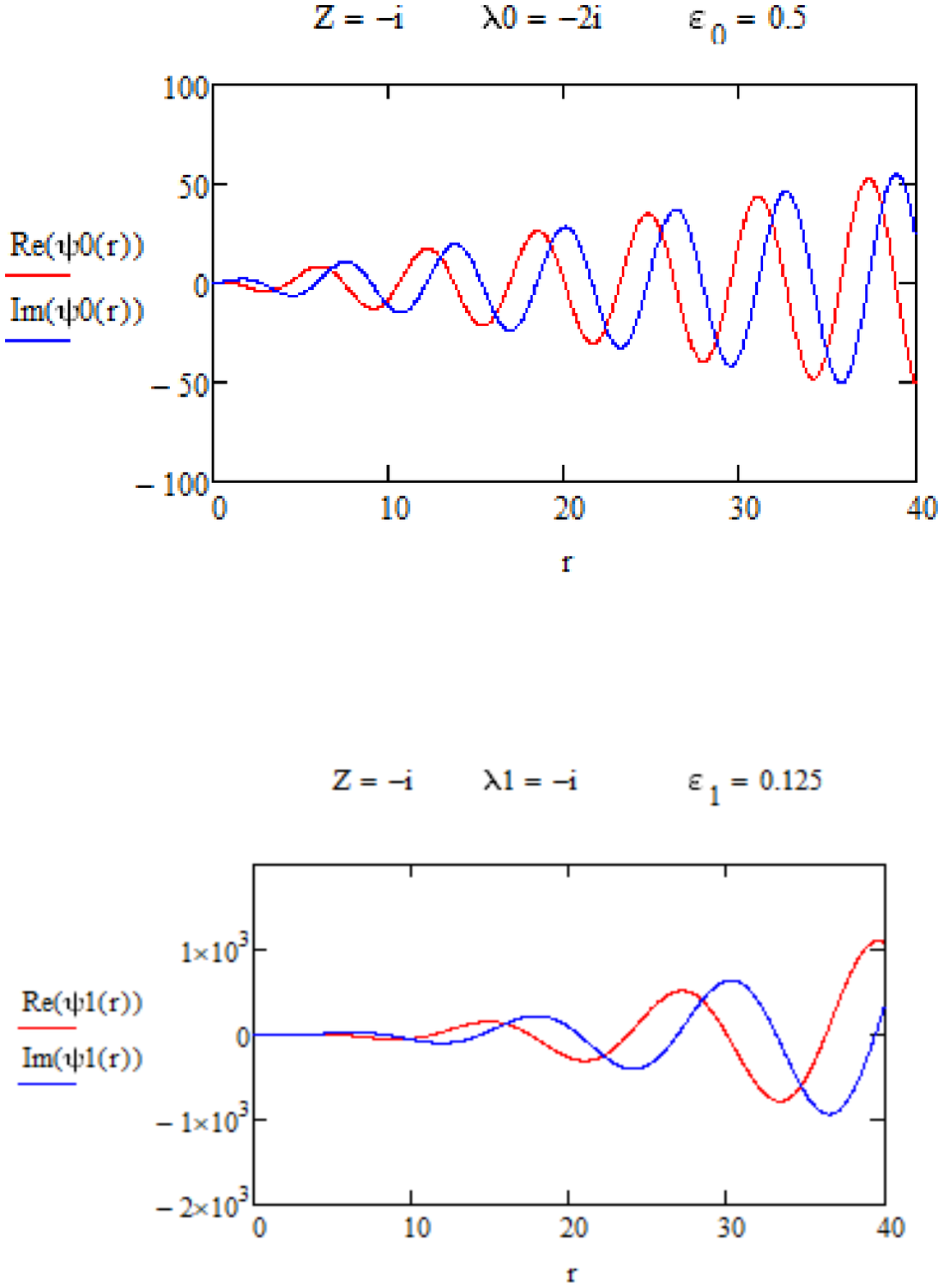}
	\caption{The real and imaginary part of the  bound-state wavefunction for the Coulomb potential with complex charge  and   different configurations shown above each plot.}
\end{figure}

 \vskip 5mm
 The radial Coulomb potential  with imaginary charge $iz$ in the atomic units $\hbar=m=1$ 
 \begin{equation}\label{coulomb}
 H= -\frac{1}{2}\frac{d^{2}}{dr^{2}}+ \frac{\ell(\ell+1)}{2r^{2}}+ \frac{iz}{r}
 \end{equation}
 The Hamiltonian \ref{coulomb} has a tridiagonal matrix representation in the non-orthogonal Laguerre basis \cite{Fishman}
 
\begin{eqnarray}
\phi_{n}(x)=B_{n}(\lambda r)^{\ell+1} e^{-\frac{\lambda r}{2}} L^{2\ell+1}_{n}(\lambda r) , \; \; n=0,1,2\dots \\ \nonumber
B_{n}= \sqrt{\frac{\lambda n!}{\Gamma (n+2\ell+ 2)}}
\end{eqnarray} 
Here $\lambda$ is a scale  parameter. If we define the $J$-matrix as $J=(H-\varepsilon)$ , then the only non-zero elements of $J$ are 
\begin{eqnarray}
(J)_{n,n}= \langle \langle \phi_{n}|H-\varepsilon|\phi_{n} \rangle \rangle = i \lambda z- (\varepsilon -\frac{\lambda^{2}}{8})(2n+2\ell+2)\\
(J)_{n,n+1}= \langle \langle 
\phi_{n}|H-\varepsilon| \phi_{n+1}\rangle \rangle = (\varepsilon + \frac{\lambda^{2}}{8}) \sqrt{(n+1) (2n+2\ell +2)}
\end{eqnarray}
The discrete spectrum can be deduced by imposing the conditions given in \cite{haidari}for specific cases,which apply here, and by the procedure explained in \cite{Yamani2} in the general case. Here, we first find
the result of demanding that $J_{\mu,\mu+ 1} = 0 $ and then use the result in the requirement that $J_{\mu,\mu} = 0 $. These conditions finally give 
\begin{eqnarray}
(\lambda)_{\mu}= \frac{2iz }{\mu+\ell+1}\\
(\varepsilon)_{\mu}= \frac{z^{2}}{2(\mu+ \ell+1)^{2}}
\end{eqnarray} 
From these results we notice that Coulomb potential with imaginary charge has  positive energy spectrum. Furthermore we notice that  there are an infinite number of positive energy bound-states in the interval
$ 0 \leq (\varepsilon)_{\mu}\leq \frac{z^{2}}{2( \ell+1)^{2}}$ . Also the wave-function associate with these bound states exhibit sinusoidal behavior as shown in figure \ref{sin}. Remarkably,  the bound-states exist for both the system with positive imaginary charge $+iz$ as well
as  for negative imaginary charge $-iz$ . This stands in contrast to usual Coulomb
Hamiltonian which support bound states only for the case that the charge $z$ is negative.  However it is not  $\mathcal{PT}$-symmetric  so we did not pursue  this direction. 

\vskip 5mm 

 For one-dimensional complex potentials the term  $ \frac{\ell (\ell+1)}{2 r^{2}}$   is absent. We apply the J-matrix
 method for the case of $\mathcal{PT}$-symmetric Rosen-Morse II potential \cite{Levai}which has   applications in optics  as shown in \cite{Roychoudhury}. The $\mathcal{PT}$-symmetric Rosen-Morse II potential is 
 
 \begin{equation}
   V(x)=2iB \tanh\delta x-A(A+ \frac{\delta \hbar}{\sqrt{2m}})\; \mathrm{sech}^{2}\delta x
   \end{equation}
   where $\alpha$ is a free real  parameter, $A,B$ are coefficients and $x$ is the coordinate variable in one-dimension. The corresponding associated energy eigenvalues are \begin{equation}E_{n}= -(A-\frac{n \hbar \delta}{\sqrt{2m}})^{2}+ \frac{B^{2}}{(A-\frac{n\hbar \delta
   	}{\sqrt{2m}})^{2}}\end{equation} 
   
   which are real. The basis function for this potential can be written in virtue of Jacobi
   polynomials $P^{\mu,\nu}_{n}$ with complex parameters since $\mu, \nu\in \mathbb{C}$ such that $\mu=\overline{\nu}$
   \begin{equation}
   \psi_{n}(y)= C_{n} (1-y)^{\alpha}(1+y)^{\beta} P^{\mu,\nu}_{n}(y)
   \end{equation}
   
   where $y=\tanh\delta x$ and The parameters  $\mu= s-n+a$, $\nu=s-n-a$  where $s= \frac{\sqrt{2m}A}{\hbar \delta}$, $\lambda= i \frac{\sqrt{2m}B}{\hbar \delta}$ and $a= \frac{\sqrt{2m}\lambda}{\hbar \delta(s-n)}$ are complex conjugate $\overline{\mu}=\nu$ unlike the real case \cite{Levai,haidari,Ismail}.
    Analytical solutions are obtainable only for the following cases $(\alpha,\beta)= (\frac{s-n+a}{2},\frac{s-n-a}{2}), (\frac{s-n+a+1}{2},\frac{s-n-a}{2}), (\frac{s-n+a}{2}, \frac{s-n-a+1}{2})$ \cite{haidari}.

  The  generalized Rosen-Morse II 
   \begin{equation}
    V_{g}(x)=2iB \tanh\alpha x- A(A+ \frac{\alpha\hbar}{\sqrt{2m}})\mathrm{sech}^{2}\alpha x+ i C \tanh \alpha x\;  \mathrm{sech}^{2} \alpha x \end{equation}
    It  has tridiagonal representation in Jacobi polynomial basis $\psi_{n}(x)=(1-\tanh(\alpha x))^{\frac{\mu}{2}} (1+\tanh(\alpha x))^{\frac{\nu}{2}} P^{(\mu,\nu)}_{n}(\tanh(\alpha x))$ in the case $(\alpha,\beta)=(\frac{s-n+a}{2},\frac{s-n-a}{2})$ where $2iB= (\frac{\delta \mu}{2})^- (\frac{\delta \nu}{2})^{2}$. \\

   Another interesting example is the $\mathcal{PT}$-symmetric Scarf II potential $V_{Scarf}(x)= -V_{1}\; \mathrm{sech}^{2} x+iV_{2}\; \mathrm{sech}x\tanh x$ for $V_{1}>0$ and $V_{2}\neq 0$ which was found to host an  exceptional point in its energy spectrum  (zero-width resonance or spectral singularity in other references ) for some wavefunction \cite{Ahmed,Ahmed1, Quesna} . Similar to the complex Rosen-Morse II potential, the complex  Scarf II potential  can be solved exactly in a Jacobi polynomial basis \cite{Quesna,Levai4}.  
	
	\section{ Pseudo-Hermitian Supersymmetric  Hamiltonians in Tridiagonal Representation }\label{2}
	We assume the matrix representation of a given $\mathcal{PT}$-symmetric  Hamiltonian  $H=BA$ in a complete orthonormal basis $\{|\phi_{n}\rangle\}$ where $n=0,1,2,\dots $, to be tridiagonal in the following manner \begin{equation}\label{tri}
	\langle\langle \phi_{n}| H| \phi_{m}\rangle\rangle=b_{n-1}\;\delta_{n,m+1}+a_{n}\; \delta_{n,m}+ b^{\star}_{n}\;\delta_{n,m-1},
	\end{equation}
	where $\{a_{n},b_{n}\}$ are complex coefficients, and $b_{-1}=0$. 
	We define the operator $A$ by its action on each element of the basis $\{|\phi_{n}\rangle\}$ as 
	\begin{equation}\label{A}
	A\;|\phi_{n}\rangle= c_{n}\;|\phi_{n}\rangle+d_{n}\;|\phi_{n-1}\rangle. 
	\end{equation}
	where the coefficients $\{c_{n},d_{n}\}$ are to be defined later expect to $d_{0}=0$. The action of operator $B$ on a given basis $\{|\phi_{n}\rangle\}$ is 
	\begin{equation}\label{Adagger}
B\;|\phi_{n}\rangle= c^{\star}_{n}\; |\phi_{n}\rangle+d^{\star}_{n+1}\; |\phi_{n+1}\rangle
	\end{equation} 
	$n=0,1,\dots $, and  $\star$ is  the complex conjugation operation.
	Note that when $c_{n}=0$ in equations \ref{A} and \ref{Adagger}, the operators $A$ and $B$ become the lowering and raising operators, respectively. 
	The product operator $BA$ has the following tridiagonal representation in the given basis $\{|\phi_{n}\rangle\}$ as follows
	\begin{align}
		\langle \langle \phi_{n} | BA|\phi_{m}\rangle \rangle= c_{m} d^{\star}_{m+1} \;\delta_{n,m+1}+ (|c_{m}|^{2}+ |d_{m}|^{2})\; \delta_{m,n} \\ \nonumber+ d_{m}c^{\star}_{m-1}\; \delta_{n,m-1}. 
	\end{align}
	where $n,m=1,2,\dots$.
	We prove this last result by acting with  \ref{A} and \ref{Adagger} on a given basis
	$\{|\phi_{n}\rangle\}$, we obtain 
	\begin{align}
		BA |\phi_{m}\rangle = B ( A |\phi_{m}\rangle) = B ( c_{m}\; |\phi_{m}\rangle + d_{m}\; |\phi_{m-1}\rangle) \\ \nonumber = 
		c_{m} c^{\star}_{m}\;|\phi_{m}\rangle + c_{m} d^{\star}_{m+1} \;|\phi_{m+1}\rangle + d_{m} c^{\star}_{m-1}\; |\phi_{m-1}\rangle+ d_{m} d^{\star}_{m} \;|\phi_{m}\rangle \\ \nonumber = 
		c_{m} d^{\star}_{m+1} \; |\phi_{m+1}\rangle + (|c_{m}|^{2}+ |d_{m}|^{2})\; |\phi_{m}\rangle+ d_{m}c^{\star}_{m-1}\; |\phi_{m-1}\rangle
	\end{align}
	Next, the matrix elements $\langle \langle \phi_{m}| BA| \phi_{n}\rangle \rangle$ can be calculated by taking the inner product of the previous relation alongside with the orthonormality condition $\langle \langle\phi_{m}|\phi_{n}\rangle\rangle= \langle \phi_{m}|\mathcal{P}|\phi_{n}\rangle=\delta_{m,n}$.   
	This gives  the desired result 
	\begin{align}\label{AA}
	\langle	\langle \phi_{m}| BA| \phi_{n}\rangle \rangle=   c_{n} d^{\star}_{n+1}\; \delta_{m,n+1}+ (|c_{n}|^{2}+ |d_{n}|^{2})\; \delta_{m,n}+  d_{n}c^{\star}_{n-1}\; \delta_{m,n-1}.
	\end{align}
	
	The coefficients $\{a_{n},b_{n}\}$ defined in \ref{tri} are connected with $\{c_{n},d_{n}\}$ via the relations 
	\begin{eqnarray}\label{coefficients1}
	a_{n}= |c_{n}|^{2}+ |d_{n}|^{2}; \\ 
	b_{n}= c_{n} \;d^{\star}_{n+1}\label{coefficients2}
	\end{eqnarray}
	From \ref{coefficients1} and \ref{coefficients2} , we find the following condition between the coefficients 
	\begin{equation}
	a_{0}d^{\star}_{1}= |b_{0}|^{2}. 
	\end{equation} 
	Now, let us calculate the supersymmetric partner Hamiltonian $H^{+}$ in the same basis $\{|\phi_{m}\rangle\}$  of the original Hamiltonian. It is not difficult to notice that $H^{+}$ has a tridiagonal representation in  the same basis  of $H$ as follows
	\begin{equation}\label{susytri}
	\langle \langle\phi_{m}| H^{+}| \phi_{n}\rangle\rangle=b^{+}_{n-1}\;\delta_{n,m+1}+a^{+}_{n}\; \delta_{n,m}+ (b^{\star}_{n})^{+}\;\delta_{n,m-1},
	\end{equation}
	
	If the operators $A$ and $B$ act on the basis elements $\{|\phi_{n}\rangle\}$ according to  \ref{A} and \ref{Adagger} respectively. The product operator $AB$ has the following tridiagonal representation in the given basis $\{|\phi_{n}\rangle\}$ as follows:
	\begin{align}\label{susytri1}
		\langle\langle \phi_{n} | A B|\phi_{m}\rangle \rangle= d_{m+1}^{\star} c_{m+1} \; \delta_{n,m+1} + ( |c_{m}|^{2}+ |d_{m+1}|^{2}) \; \delta_{m,n}+ \\ \nonumber c^{\star}_{m} d_{m} \; \delta_{n,m-1} . 
	\end{align}
	where $n,m=1,2,\dots$. 
	The coefficients in \ref{susytri} are connected with $c_{n}$ and $d_{n}$ in the following way 
	\begin{eqnarray}\label{a}
	a_{n}^{+}=  |c_{n}|^{2}+ |d_{n+1}|^{2}, \\
	b_{n}^{+}= d^{\star}_{n+1}c_{n+1}. \label{b}
	\end{eqnarray}
	If we consider $A$ and $B$ as lowering and raising operators respectively, we would then have vanishing $c_{n}$ and $c^{\star}_{n}$  coefficients. In that case, both $H$ and $H^{(+)}$ are diagonal in this basis   $\{|\phi_{n}\rangle\}$ as follows 
	\begin{eqnarray}
	H_{nm}= |d_{n}|^{2} \delta_{m,n}, \\ \nonumber
	H^{(+)}_{nm} = |d_{n+1}|^{2} \delta_{m,n}. 
	\end{eqnarray}
	As a final comment, the energy eigenvalues $E^{(+)}_{n} $ of the  superpartner Hamiltonian $H^{(+)} $ are connected with the energy eigenvalues $E_{n}$ of the Hamiltonian $H$ through the relation $E^{(+)}_{n}=E_{n+1}$  as expected for any SUSY quantum mechanical model. 
	
	\section{Eigenstates expansion using orthogonal polynomials }
	The action of  Hamiltonian $H$ in the tridiagonal representation is 
	\begin{equation}\label{tridiagonal}
	H|\phi_{n}\rangle= b_{n-1}|\phi_{n-1}\rangle+ a_{n} |\phi_{n}\rangle+ b^{\star}_{n} |\phi_{n+1}\rangle
	\end{equation} 
	$n=0,1,\dots$. One other hand, the Hamiltonian $H$ obeys the following eigenvalue equation 
	\begin{equation}
	H|\psi_{E}\rangle= E | \psi_{E}\rangle,
	\end{equation}
	where $|\psi_{E}\rangle$ is the set of energy eigenstates. It is convenient in our case to expand $|\psi_{E}\rangle$ in term of the basis set $\{|\phi_{n}\rangle\}$ as 
	\begin{equation}\label{expansion}
	|\psi_{E}\rangle= \Sigma_{n=0}^{\infty}\;  C_{n}(E) \; |\phi_{n}\rangle. 
	\end{equation}
	Here $C_{n}(E)$ are defined as complex  coefficients in general. 
	Then, making use of equation \ref{expansion} and the orthonormality of basis set $\{|\phi_{n}\}$ in \ref{tridiagonal} gives the following recurrence relation of the expansion coefficients $C(E)$
	\begin{equation}\label{polynomial}
	E \;C_{n}(E)= b_{n-1}\; C_{n-1}(E) + a_{n}\; C_{n}(E)+ b^{\star}_{n}\; C_{n+1}(E).  
	\end{equation}
	The coefficients $C_{n}(E)$ satisfy the following  orthonormality condition in case of discrete energy spectrum 
	\begin{align}
		\int _{\Omega (E)} C^{\star}_{n}(E) \; C_{m} (E) \; dE= \int _{\Omega(E)} \langle\langle \phi_{n}| \psi_{E}\rangle \rangle \langle\langle\psi_{E}|\phi_{m}\rangle\rangle \; dE=\langle \langle \phi_{n}|\phi_{m}\rangle\rangle= \delta_{n,m}. 
	\end{align}
	where $\Omega(E)$ is the energy support of the Hamiltonian $H$. For general case, when we have both discrete and continuous energy spectrum, the orthonormality condition becomes 
	\begin{equation}
	\delta_{m,n}= \Sigma_{\alpha}\; C_{\alpha}^{\star} (E_{\alpha})\; C_{m}(E_{\alpha})+ \int _{\Omega (E)} C^{\star}_{n}(E) \; C_{m} (E) \; dE.
	\end{equation} Let us define $P(E)$ as 
	\begin{equation} \label{P}
	P_{n}(E)= \frac{C_{n}(E)}{C_{0}(E)}
	\end{equation}
	with $P_{0}(E)=1$ and $P_{1}(E)= (E-a_{0}) (b^{\star}_{0})^{-1}$. 
	By plugging \ref{P} in \ref{polynomial} we obtain the following three-recursion formula 
	\begin{equation}
	E P_{n}(E)= b_{n-1} P_{n-1}(E)+ a_{n} P_{n} (E) +b^{\star}_{n} P_{n+1}(E). 
	\end{equation}
	The coefficients $(c_{n}, d_{n})$ are related to the values at zero of consecutive polynomial $P_{n}(0)$ as 
	\begin{eqnarray}
	(d^{\star}_{n+1})^{2}= -b_{n}\; \frac{P_{n}(0)}{P_{n+1}(0)},   \\
	|c_{n}|^{2}= -b^{\star}_{n}\; \frac{P_{n+1}(0)}{P_{n}(0)}. 
	\end{eqnarray}
	Since the supersymmetric partner Hamiltonian has the same tridiagonal representation in the same basis, we write
	\begin{equation}
	E P_{n}^{(+)}(E)= b_{n-1}^{(+)}\; P_{n-1}^{(+)}+ a^{(+)}_{n}\; \
	P^{(+)}_{n}(E)+ (b^{(+)}_{n})^{\star}\; P^{(+)}_{n+1}(E). 
	\end{equation}
	With the initial conditions $ P^{(+)}_{0}(E)=1$ and $ P^{(+)}_{1}(E)= \frac{(E-a_{0}^{+})}{ ({b}_{0}^{(+)})^{\star}}$. The polynomials  $P_{n}(E)$ are related to their supersymmetric partner $P^{(+)}_{n}(E)$  via the following relation 
	\begin{equation}\label{kernel}
	P^{(+)}_{n}(E)=\sqrt{\frac{b_{0}P_{1}(0)}{b_{n}P_{n}(0)P_{n+1}(0)}}\; K_{0}(E,0), 
	\end{equation}
	where $K_{n}(E,0)= \Sigma_{j=0}^{n} P_{j}(E) P_{j}(0)$ denotes the kernel polynomials \cite{Yamani}.

	\section{Supersymmetry of non-Hermitian  $\mathcal{PT}$-symmetric  Morse oscillator}
	The generalized non-Hermitian $\mathcal{PT}$-symmetric  Morse potential is \cite{Morse,Znojil,Quesna1,Quesna2}
	
	\begin{eqnarray}\label{generalized}
	V(x)= V_{1} e^{-2i \alpha x}-V_{2} e^{-i\alpha x}\\ 
	= V (e^{-2i\alpha x}- q e^{-i\alpha x}). 
	\end{eqnarray}
	where $V=V_{1}$ and $q=\frac{V_{2}}{V_{1}}$. \\
	We choose $V=V_{0}$ and $q=2$ , then the potential reads  
	\begin{equation}\label{morsepotential}
	V(x)= V_{0} (e^{-2i\alpha x} -2e^{-i \alpha x}), 
	\end{equation} 
	where $V_{0}$ and $\alpha$ are real parameters of the potential.   One interesting feature of  the Morse Hamiltonian is the possibility of having a discrete energy spectrum depending on the parameter $V_{0}$ besides its continuous ones. However, in contrast to real case, the potential \ref{morsepotential} is bounded and  periodic function for fixed values of $V_{0}$. The total Hamiltonian $H$ is unbounded non-self adjoint operator. Moreover, it  has imaginary and real parts.   Another alternative way for obtaining real eigenvalues of energy was addressed in   \cite{Znojil} for generalized Morse potential defined in \ref{generalized} .
	We apply our developed formalism in earlier sections to the case of  $\mathcal{PT}$-symmetric shifted Morse Hamiltonian written in atomic units $\hbar=m=1$ as \cite{Yamani2}
	\begin{equation}
	H= -\frac{1}{2}\; \frac{d^{2}}{dx^{2}}+ V_{0}\;(e^{-2i\alpha x} -2e^{-i \alpha x}) - \frac{1}{2}\alpha^{2} D^{2}, 
	\end{equation}
	where $D= \sqrt{\frac{-2V_{0}}{\alpha^{2}}}-\frac{1}{2}$. \vskip 5mm Before we start our analysis   it would be important to trace  the energy spectrum of the $\mathcal{PT}$-symmetric Morse potential and verify it is real on some interval. To answer this exactly we numerically find the eigenvalues of \ref{morsepotential}  in a Hermite orthogonal basis endowed with a  scale parameter $\lambda$ 
	\begin{equation}
	\psi_{n}(x)= A_{n} e^{-\frac{\lambda^{2}x^{2}}{2}} H_{n}(\lambda x), 
	\end{equation}
	where $A_{n}= \sqrt{\frac{\lambda}{2^{n}n!\sqrt{\pi}}}$. This basis does not make \ref{morsepotential} tridiagonal but it is nice on two counts. First it is complete in the $x$-space over the interval $(-\infty,\infty)$ which is the range of Morse potential. Second, for any desired value $N$, it generates two sets of numbers , the discrete points $\{x_{\mu}\}_{\mu}^{N-1}$ and the associated matrix $\{\Gamma_{n,\mu}\}_{\mu,n}^{N-1}$ to enable us doing integrals approximately, but accurately, simply as a quadrature 
	\begin{equation}
	S_{n,m}\equiv\int dx\;  \psi_{n}(x) S(x) \psi_{n}(x) \simeq \Sigma_{\mu=0}^{N-1}\;  \Gamma_{n,\mu} \; S(x_{\mu}) \Gamma_{m,\mu
	}. 
	\end{equation}
	For any function $S(x)$ such as $\mathcal{PT}$-symmetric Morse potential. For $N=70$ and $\lambda=12$ we were able to verify the reality of eigenvalues for the $\mathcal{PT}$-symmetric Morse potential. For lower scale parameter
	some complex eigenvalues creep usually in pairs as shown in Figure 1 and  2. If we go higher than $\lambda=10$, we have a pure real spectrum. 	The Gauss quadrature treatment can be used for all complex potentials. It is convenient in this case to consider orthogonal  polynomials with three-recurrence relation on the interval of study\cite{Golub,Broad,Reinhardt}.
	
	\begin{figure}\label{figure1}
		\includegraphics[width=8cm]{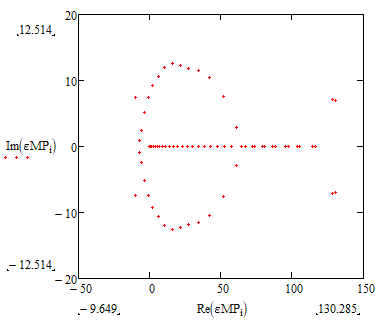}
		\caption{The imaginary part of the $\mathcal{PT}$-symmetric Morse potential with respect to the real part with $\lambda=1.5$ and $N=70$. We notice that complex (also the real)  eigenvalues creep usually in pairs.}
	\end{figure}
	
		\begin{figure}\label{figure2}
		\includegraphics[width=8cm]{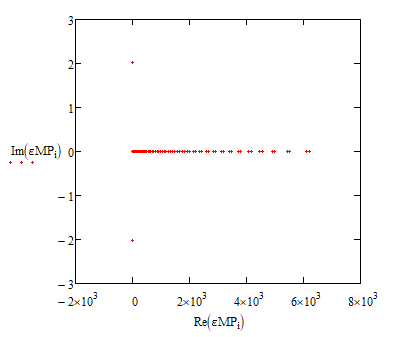}
		\caption{The imaginary part of the $\mathcal{PT}$-symmetric Morse potential with respect to the real part with $\lambda=10$ and $N=70$. We get a totally
			real spectrum except for one pair of complex energy eigenvalues. If we jack up the value of scale parameter higher than $\lambda=10$ this last pair disappear, and we have a pure real spectrum. }
	\end{figure}
  \vskip 5mm
	The  complex Morse Hamiltonian has tridiagonal representation using the following orthogonal basis
	\begin{equation}\label{basis}
	\phi_{n}(x)= \sqrt{\frac{ i \alpha n! }{\Gamma(n+2\gamma+1)}} \; (\xi)^{\gamma+\frac{1}{2}} \; e^{-\frac{1}{2}\xi}\; L_{n}^{2\gamma}(\xi), 
	\end{equation}
	where $\xi=   \sqrt{\frac{-8V_{0}}{\alpha^{2}}}\; e^{-i\alpha x}$ and $L_{n}^{2\gamma}(\xi)$ is the associated Laguerre polynomial\cite{MOS}. The parameter $\gamma$ can take any value except    $-\frac{1}{2}$. In contrast to the real case,  $L_{n}^{\mu}(z)$ is a function of  complex variable $z$. The basis \ref{basis}  vanishes asymptotically when $x\rightarrow \pm \infty$ if and only if $V_{0}$ is negative \cite{Levai1}. We considered one set of basis \ref{basis} for simplicity. However as emphasized in \cite{Znojil} it is possible to  write the total wavefunction for the one-dimensional  $\mathcal{PT}$-symmetric Morse potential as a sum of two terms one of them is what I considered in \ref{basis} and other term which is the same but with $\gamma\rightarrow -\gamma$.  The basis set $\ref{basis}$ can be written in term of confluent hypergeometric function of the first kind $1F_{1}(a;b;x)$ since $L^{2\gamma}_{n}(\xi)= \frac{(2\gamma+1)_{n}}{n!} 1F_{1}(-n; 2\gamma+1;\xi)$ where $(2\gamma+1)_{n}$ is the Pochhammer symbol \cite{KS}.

	The matrix representation of the shifted Morse Hamiltonian is 
	\begin{equation}
	\langle \langle \phi_{n} |H|\phi_{m}\rangle \rangle= b_{n-1}\; \delta_{n,m+1}+ a_{n}\; \delta_{n,m} +  b^{\star}_{n}\;  \delta_{n,m-1}. 
	\end{equation}
	Since the Hamiltonian $H$ obeys the eigenvalue equation $H|\psi_{E}\rangle= E | \psi_{E}\rangle$ we can make the following expansion $|\psi_{E}\rangle= \Sigma_{n=0}^{\infty}\;  C_{n}(E) \; |\phi_{n}\rangle$. We note that $C
	(E)$ satisfies a recursion relation similar to \ref{polynomial}. More explicitly, we have the following 
	\begin{align}
		E\; C_{n}(E)=   -\frac{\alpha^{2}}{2}\; \sqrt{n(n+2\gamma)} \big(n-\frac{1}{2}+ \gamma-D)\; C_{n-1}(E)   
		\\ \nonumber	+ \frac{\alpha^{2}}{2}\; \bigg(\big(n+\gamma+\frac{1}{2}- D\big)^{2}  + n(n+2\gamma)- D^{2}\bigg)\; C_{n}(E)\\ \nonumber 
		-\frac{\alpha^{2}}{2} \sqrt{(n+1) (n+2\gamma+1)} \big(n+\gamma+ \frac{1}{2}-D\big)\; C_{n+1}(E).&&
	\end{align}
	In contrast to the real case, $D$ is imaginary for positive $V_{0}$. 
	The polynomials $\mathcal{C}_{n}(E)$ can be defined in terms of the continuous dual Hahn polynomials $S_{n}$ as 
	\begin{eqnarray}
	C_{n}(E)= \sqrt{\frac{\Gamma(n+2\gamma+1)}{n!}}\; S_{n}(\lambda^{2}; -D, \gamma+\frac{1}{2}, \gamma+\frac{1}{2})\\ \nonumber = \sqrt{\frac{\Gamma(n+2\gamma+1)}{n!}}\; \lim_{d\rightarrow \infty} \frac{W_{n}(\lambda^{2}; -D, \gamma+\frac{1}{2}, \gamma+\frac{1}{2},d)}{(-D+d)_{n}},  
	\end{eqnarray} 
where 	$\lambda=\sqrt{-2\alpha^{-2}E-D^{2}}$ and $W_{n}$ is the Wilson polynomials with  $d$ being  an arbitrary parameter \cite{KS}.    From the previous relation we obtain  the following exact expression for $P_{n}$ , 
	\begin{align}\label{pn}
		P_{n}\left( E \right)= \frac{C_{n}(E)}{C_{0}(E)} \\ \nonumber =\frac{\left( -1\right) ^{n}\left(
			\gamma +\frac{1}{2}-D)\right) _{n}}{\sqrt{n!\left( 2\gamma +1\right) _{n}}} \times \\ \nonumber 
		\text{ }_{3}\digamma _{2}\left(
		\begin{array}{ccc}
			-n & 1-D+i\lambda & 1-D-i\lambda \\
			& \gamma +\frac{1}{2}-D & \gamma +\frac{1}{2}-D
		\end{array}
		\mid 1\right)  
	\end{align}
	Following the developed formalism in \ref{2}, we need to calculate the coefficients $(c_{n}, d_{n})$ in order to find $a^{+}_{n}$ and $b^{+}_{n}$ that appear in the matrix representation of  supersymmetric partner Hamiltonian $H^{+}$ \ref{susytri}. Thus, we have  the following relations 
	\begin{equation}\label{c1}
	c_{n}= \frac{i \alpha}{\sqrt{2}}(n+\gamma+\frac{1}{2}-D),
	\end{equation}
	
	\begin{eqnarray}\label{dn}
	d_{n+1}= -\frac{i \alpha}{\sqrt{2}}\; \sqrt{(n+1) (n+1+2\gamma)}.
	\end{eqnarray}
	
	By plugging \ref{c1}, \ref{dn} in \ref{a} and \ref{b}, we find 
	\begin{equation}
	a_{n}^{\left( +\right) }=\frac{\alpha ^{2}}{2}\left( \left( n+1\right)\left( n+2\gamma+1 \right)\right) +\big(n+\gamma+\frac{1}{2}- D\big)^{2}
	\end{equation}
	\begin{equation}
	\mathit{\ }b_{n}^{\left( +\right) }=-\frac{\alpha ^{2}}{2}\sqrt{\left(
		n+1\right) \left( n+2\gamma+1 \right) }\left( n+\gamma +\frac{3}{2}-D\right)
	\end{equation}
	for every $n=0,1,2,...$ . From \ref{kernel} we can find the relation for $P_{n}^{(+) }\left( E \right)$ provided that we have  determined $P_{n}\left( E \right)$ in \ref{pn} 
	\begin{align}\label{pn+}
		P_{n}^{(+) }\left( E \right) =\frac{\left(
			-1\right) ^{n}\left( \gamma +\frac{3}{2}-D)\right) _{n}}{\sqrt{n!\left(
				2\gamma +2\right) _{n}}} \times \\ \nonumber \text{ }_{3}\digamma_{2}\left(
		\begin{array}{ccc}
			-n & 1-D+i\lambda & 1-D-i\lambda \\
			& \gamma +\frac{3}{2}-D & \gamma +\frac{3}{2}-D
		\end{array}
		\mid 1\right)  .
	\end{align}
	comparing $\ref{pn+}$ with $\ref{pn}$ we found that the supersymmetric partner of $P_{n}(E)$ can be obtained by the substitution $\gamma\rightarrow \gamma+1$ as expected for supersymmetric theory. 
	\section{Conclusion}
In this work,	we have extended the study of supersymmetric  tridiagonal Hamiltonians to include  $\mathcal{PT}$-symmetric non-Hermitian Hamiltonians.  Our developed formalism can be used for  both Hermitian and  non-Hermitian Hamiltonians so it provides  a generalization to \cite{Yamani}. Moreover, we developed  the notion of $J$-matrix method for complex potentials.  We have calculated the matrix coefficients  for the Hamiltonian $H$ and its supersymmetric partner $H^{(+)}$. In contrast to the Hermitian Hamiltonian case described in \cite{Yamani}, we found the coefficients $a_{n}$ and $b_{n}$ in eq. \ref{tri} to be  imaginary. Moreover we have expanded the eigenstates of  $H$ and its supersymmetric partner in terms of orthogonal polynomials. We proved that orthogonal polynomials attached to $H^{(+)}$ can be determined from the orthogonal polynomials attached to $H$ with the help of kernel polynomials.  We examined our formalism by solving the $\mathcal{PT}$-symmetric shifted Morse potential exactly. In this case, we found the coefficient polynomials $P_{n}(E)$ and its supersymmetric partner $P^{+}_{n}(E)$ to be given in term of the hypergeometric function $ \text{ }_{3}\digamma_{2}$. These results make our task of classifying complex potentials with real energy eigenvalues easier when one consider the Askey scheme which is based mainly  hypergeometric functions \cite{KS}. \vskip 5mm

{\bf Acknowledgment.}	The author is  very much indebted to Professor  Miloslav  Znojil  for his comments on earlier draft  of this work.  The author would like to thank
the anonymous Referees for the constructive comments which improved the presentation of this work. 
	\begin{footnotesize}

	\end{footnotesize}
	

\begin{thebibliography}{9}
			\bibitem{Witten}
			E. Witten, \textit{Dynamical breaking of supersymmetry}, Nucl. Phys. B {\bf 188} 513 (1981).
			\bibitem{Krive}
			L.E. Gendenshtein and I.V. Krive, \textit{Supersymmetry in Quantum Mechanics
			}, Sov.Phys.Usp. {\bf 28} 645-666 (1985), Usp.Fiz.Nauk {\bf 146}  553-590 (1985). 
			\bibitem{Haymaker}
			R.W. Haymaker and A.R.P. Rau, \textit{Supersymmetry in quantum mechanics}, American Journal of Physics {\bf 54}, 928 (1986).
			\bibitem{Cooper2001} F. Cooper, A. Khare, and U.P. Sukhtame, \textit{Supersymmetry in Quantum Mechanics}, World
			Scientific, Singapore (2001).
			\bibitem{Bagchi}
			B.K. Bagchi, \textit{Supersymmetry in quantum and classical mechanics}
			, Chapman and  Hall/CRC monographs and surveys in pure and applied mathematics 116 (2001).
			\bibitem{Infeld}
			L. Infeld and T.E. Hull, \textit{The factorization method} , Review of Modern Physics {\bf23} (1951) 21.
			\bibitem{Dong} 
			Shi-Hai Dong, 
			\textit{Factorization Method in Quantum Mechanics}
			, Springer (2007). 
			
		\bibitem{Gend} L. Gendenshtein, \textit{Derivation of Exact Spectra of the Schrodinger Equation by Means of Supersymmetry
		} JETP Lett.\textbf{38}  356 (1983). 
				
					
			\bibitem{Fishman}H.A. Yamani and L. Fishman, \textit{J-matrix method:Extensions to arbitrary angular momentum and to Coulomb scattering}, Journal of Mathematical Physics {\bf 16}, 410 (1975). 
			\bibitem{haidari}
			A.D. Alhaidari, \textit{An extended class of $L^{2}$-series solutions of the wave equation}, Annals of Physics  {\bf 317}, 152-174 (2005). 
			
		
			
			\bibitem{Yamani} H.A. Yamani and Z. Mouayn, \textit{Supersymmetry of tridiagonal Hamiltonians}, J. Phys. A: Math. Theor. \textbf{47}  265203 (2014).
			 
			\bibitem{Yamani2}
			H.A. Yamani and Z. Mouayn, \textit{Supersymmetry of the Morse oscillator }, Reports on Mathematical Physics \textbf{78} No.3 (2016).
			\bibitem{Bessis}
			 Daniel Bessis, unpublished (1992).
			\bibitem{Bender}
			C.M. Bender and S. Boettcher, \textit{Real Spectra in Non-Hermitian Hamiltonians Having PT-Symmetry}, Physics Review Letters {\bf 80}   5243 (1998) .
			\bibitem{jones}
			C.M. Bender, D.C. Brody, and H.F. Jones, \textit{Complex Extension of Quantum Mechanics} , Phys. Rev. Lett. {\bf 89}, 270401 (2002). 
			\bibitem{Bender1}
			C.M. Bender, \textit{Making Sense of Non-Hermitian Hamiltonians
			},Rept.Prog.Phys. {\bf 70}:947,(2007). 
			\bibitem{Bender2}
			Carl M. Bender, \textit{PT Symmetry
				In Quantum and Classical Physics}, World Scientific Publishing Europe Ltd (2018).
				\bibitem{Optic}
			R. El-Ganainy, K.G. Makris, D.N. Christodoulides, and Z.H. Musslimani, \textit{Theory of coupled optical PT-symmetric structures
			}, Optics Letters  Vol. {\bf 32}, Issue 17, pp. 2632-2634 (2007).
		\bibitem{Musslimani}
		K.G. Makris, R. El-Ganainy, D.N. Christodoulides, and Z.H. Musslimani, \textit{Beam Dynamics in $\mathcal{PT}$	Symmetric Optical Lattices}, Phys. Rev. Lett.{\bf 100} , 103904 (2008). 
		
		\bibitem{Ramezani} 
		H. Ramezani, T. Kottos, R. El-Ganainy, and D.N. Christodoulides,\textit{Unidirectional Nonlinear PT-symmetric Optical Structures
		}, Phys. Rev. A {\bf 82}, 043803 (2010). 
	\bibitem{Longhi}
	S. Longhi, \textit{$\mathcal{PT}$-symmetric laser absorber}, Phys. Rev. A {\bf 82}, 031801(R) (2010).
	
			\bibitem{Jin}
			L. Jin, Z. Song, \textit{A physical interpretation for the non-Hermitian Hamiltonian}, J. Phys. A: Math. Theor.{\bf 44}, 375304 (2011). 
			\bibitem{Jin1}
			S.M. Zhang, X.Z. Zhang, L. Jin, and Z. Song,\textit{High-order exceptional points in supersymmetric arrays},  Phys. Rev. A {\bf 101}, 033820 (2020).
			\bibitem{Lin2}
			X. Z. Zhang, L. Jin, Z. Song, \textit{Dynamic magnetization in non-Hermitian quantum spin systems	}, Phys. Rev. B {\bf 101}, 224301 (2020). 
		\bibitem{Berakdar1}
		X. Wang, G. Guo and J.
		 Berakdar, \textit{Steering magnonic dynamics and permeability at exceptional points in a parity-time symmetric waveguide},  Nature Communications {\bf 11},  5663 (2020).
			\bibitem{Znojil1}
			M. Znojil, F. Cannata, B. Bagchi, and R. Roychoudhury, \textit{Supersymmetry without hermiticity
			}, Physics Letters B {\bf 483}, 284 (2000).
			\bibitem{Mallik}
			B. Bagchi, S. Mallik and C. Quesne, \textit{Generating Complex Potentials with Real Eigenvalues in Supersymmetric Quantum Mechanics},  Int.J.Mod.Phys. A {\bf 16}  2859-2872 (2001).
				\bibitem{Levai3}
			G. Levai, Exactly Solvable $\mathcal{PT}$-symmetric models, contribution in Carl M. Bender, \textit{PT Symmetry
				In Quantum and Classical Physics}, World Scientific Publishing Europe Ltd (2018).
			\bibitem{tri}
			F. Bagarello, F. Gargano and F.Roccati, \textit{Tridiagonality, supersymmetry and non self-adjoint Hamiltonians}, J. Phys. A: Math. Theor. {\bf 52}, 355203 (2019).  
			\bibitem{Likhtman}
			Y.A. Gel'fand and E.P. Likhtman, \textit{Extension of the Algebra of Poincare Group Generators and Violation of p Invariance}, JETP Lett.  {\bf 13}  323 (1971).
			\bibitem{Ramond}
			P. Ramond, \textit{Dual Theory for Free Fermions
			}, Phys. Rev. D {\bf  3} , 2415 (1971). 
			\bibitem{Wess}
			J. Wess  and B. Zumino, \textit{Supergauge transformations in four dimensions},  Nucl. Phys. B {\bf 70}: 39 (1974).
	
			\bibitem{Mostafazadeh}
			A. Mostafazadeh, \textit{Pseudo-Hermiticity versus PT Symmetry: The necessary condition for the reality of the spectrum  non-Hermitian Hamiltonian},  Journal of Mathematical Physics {\bf 43}, (2002).
				
			\bibitem{Sukumar}
			C.V. Sukumar, \textit{Supersymmetry, factorisation of the Schrodinger equation and a Hamiltonian hierarchy},  J. Phys. A: Math. Gen.  {\bf 18} L57-L61 (1985).
		
			 
			
			\bibitem{Bagchi1}
			B. Bagchi, S. Mallik and C. Quesne, \textit{pt-symmetric square well and the associated susy hierarchies},  Modern Physics Letters A {\bf 17}, No. 25, pp. 1651-1664 (2002).
			\bibitem{Mourad}
			M.E.H. Ismail and E. Koelink, \textit{The J -matrix method},  Advances in Applied Mathematics 46 , 379-395 (2011). 
			\bibitem{Levai1}
			G. Levai and M.Znojil, \textit{ J. Phys. A: Math. Gen.} {\bf 33}  7165–7180 (2000).
			\bibitem{Levai}
			G. Levai and E. Magyari,\textit{The PT-symmetric Rosen-Morse II potential: effects of the asymptotically non-vanishing imaginary potential component}, J. Phys. A: Math. Theor. {\bf 42} 195302 (2009). 
			\bibitem{Roychoudhury}
			B. Midya and R. Roychoudhury, \textit{Nonlinear localized modes in 
				$\mathcal{PT}$-symmetric Rosen-Morse potential wells
			}, Phys. Rev. A {\bf 87}, 045803 (2013).
			\bibitem{Ahmed}
			Z. Ahmed, \textit{Real and complex discrete eigenvalues in an exactly solvable one-dimensional complex PT invariant potential
			}, Phys.Lett.A {\bf 282} 343-348 (2001)  . 
		\bibitem{Ahmed1}
		Z. Ahmed, \textit{Addendum. Real and complex discrete eigenvalues in an exactly solvable one-dimensional complex PT invariant potential
		}, Phys.Lett.A {\bf 287} 295-296 (2001) . 
			\bibitem{Quesna}
			B. Bagchi and C. Quesne, \textit{An update on PT-symmetric complexified Scarf II potential, spectral singularities and some remarks on the rationally-extended supersymmetric partners
			}, J.Phys.A {\bf 43}:305301,(2010).
		\bibitem{Levai4}
		G. Levai, F. Cannata, and A. Ventura, \textit{PT symmetry breaking and explicit expressions for the pseudo-norm in the Scarf II potential
		}, 	Phys. Lett. A {\bf 300}  271 (2002).
			\bibitem{Ismail}
			Mourad E. H. Ismail ,\textit{ Classical and Quantum Orthogonal Polynomials in One Variable}, Cambridge University Press (2005). 
			\bibitem{Golub} 
		G.H. Golub and J.H. Welsch,  \textit{Calculation of Gauss Quadrature Rules}, Mathematics of Computation {\bf 23}  (106): 221–230 (1969). 
		\bibitem{Broad}
		J.T. Broad, \textit{Gauss quadrature generated by diagonalization of $H$ in finite $L^{2}$ bases}, Phys. Rev. A {\bf 18}, 1012 (1978).
		
			\bibitem{Reinhardt}
			W.P. Reinhardt, \textit{Universality	Properties	of	Gaussian	Quadrature,	The	Derivative	Rule,	and	a	Novel Approach	to	Stieltjes	Inversion}, available at https://arxiv.org/pdf/1812.02196.pdf
				\bibitem{Morse} 
			P.M. Morse, \textit{Diatomic Molecules According to the Wave Mechanics. II. Vibrational Levels}, Phys. Rev. {\bf 34}, 57 (1929). 
			\bibitem{Znojil} 
			M. Znojil, \textit{Exact solution for Morse oscillator in PT-symmetric quantum mechanics
			}, Physics Letters A
		 {\bf 264},  108-111 (1999) .
		 \bibitem{Quesna1} 
		 B. Bagchi and C. Quesne, \textit{sl(2, C) as a complex Lie algebra and the associated non-Hermitian Hamiltonians with real eigenvalues
		 }, 	Phys.Lett. A {\bf 273}  285-292 (2000). 
			\bibitem{Quesna2}
			B. Bagchi and C. Quesne, \textit{Non-Hermitian Hamiltonians with real and complex eigenvalues in a Lie-algebraic framework
			}, Phys.Lett.A {\bf 300}:18-26,(2002).
		
		
			
			
			
			\bibitem{MOS} W. Magnus, F. Oberhettinger, R.P. Soni, \textit{Formulas and Theorems
			for the Special Functions of Mathematical Physics}, Springer-Verlag Berlin
			Heidelberg New York, (1966).
			
			\bibitem{KS}
			R. Koekoek , R.F. Swarttouw, \textit{The Askey-scheme of hypergeometric orthogonal polynomials and its q-analogue}, Delft University of Technology, Faculty of Information Technology and Systems, 
			Department of Technical Mathematics and Informatics
			Report no. 98-17, available at  https://homepage.tudelft.nl/11r49/documents/as98.pdf
			
			
		\end{thebibliography}
\end{document}